\documentclass[12pt,thmsa,a4paper,titlepage]{article}
\usepackage{sw20lart}

\input tcilatex
\QQQ{Language}{
American English
}

\begin{document}

\title{Intrinsic Pinning in Layered Antiferromagnetic Superconductor}
\author{Tomasz Krzyszto\'{n}}
\date{Institute for Low Temperatures and Structure Research, Polish Academy of
Sciences,50-950 Wroc{\l}aw,Poland\thanks{%
E-mail krzyszto@apollo.int.pan.wroc.pl}}
\maketitle

\begin{abstract}
The intrinsic pinning in high-temperature superconductor with long range
antiferromagnetic order of rare-earth ions confined to the isolating planes
is described. The interaction of antiferromagnetic and superconducting
subsystems in the mixed state may lead to the creation of spin-flop domain
along the vortex core. It is shown that the behavior of several physical
quantities such as magnetization, activation energy, current-voltage
characteristic and flux creep changes when the direction of the external
field changes in the basal a-b plane. It is also shown that the decay of the
trapped flux is logarithmic function of time.
\end{abstract}


\section{ INTRODUCTION}

The problem of interplay between long-range magnetism and superconductivity
was studied thoroughly several years ago in Chevrel phases, rhodium borides
and rhodium stannides \cite{1}. On the contrary, this phenomenon has not
attracted much attention in the high-temperature superconductors (HTS). Both
competitive phenomena coexist in classical superconductors because 4f
electrons of rare-earth (RE) atoms responsible for magnetism and 4d
electrons of molybdenum chalcogenide clusters (rhodium boride or rhodium
stannide clusters) responsible for superconductivity are spatially separated
from each other. The situation seems to be very similar in layered HTS. Here
magnetic order is produced by the regular lattice of RE ions occupying
isolating layers electrically isolated from the superconducting Cu-O planes.
Therefore spin interaction between the local magnetic moments and the
conduction electrons is to weak to inhibit superconductivity. The typical
example of that layered system is ErBa$_2$Cu$_3$O$_7$. This compound has
tetragonal unit cell with small orthorhombic distortion in the a-b plane 
\cite{2}. The Er ions form two sublattice antiferromagnetic structure of
magnetic moments lying parallel and antiparallel to the b direction in the
a-b plane \cite{3}. Recently discovered RE nickel boride-carbides \cite{4}
may serve as an another example of layered magnetic superconductors. The
structure of these compounds is similar to that of HTS and consists of RE-C
layers separated by Ni$_2$B$_2$ sheets. For example in ErNi$_2$B$_2$C \cite
{5} the antiferromagnetic structure is associated with magnetic moments of Er%
$^{+3}$ ions, which order below 6 K in a transversely polarized planar
sinusoidal structure propagating along a or b axis with Er moments parallel
to the a or b axis respectively.

In this paper we consider the structure shown on Fig.1 that we believe
simulates real structure of many antiferromagnetic layered superconductors.
It consists of superconducting layers of thickness d$_s$ and magnetic
moments of RE ions running parallel and antiparallel to the b axis in the
isolating layers of thickness $d_i$, $d=d_s+d_i\approx d_i$.

An antiferromagnet with two sublattices shows different magnetization
behavior in an external magnetic field applied parallel or perpendicular to
the easy axis \cite{6}. The main difference is that antiferromagnetic
configuration (AF) is unstable in perpendicular field. When the external
field is applied parallel to the magnetic moments the AF phase is stable up
to the critical field $H_T$ above which spin-flop phase develops. This was
the basic idea put forward to explain strange behavior of DyMo$_6$S$_8$ in
applied magnetic field \cite{7}. Neutron scattering experiments \cite{8}
have shown that this compound develops long range AF state in the presence
of superconductivity. Unexpectedly some ferromagnetic peaks were observed in
the mixed state when the applied field exceeded 200 Oe, considerably below $%
Hc_2$. The idea of induced spin-flop transition near the core of the vortex
was introduced to explain these observations. The ferromagnetic-like order
confined along the vortices was belived to be responsible for additional
peaks above 200 Oe.

In the present model (see also \cite{9}) we come across two following
situations:

1. The external magnetic field is pointing in the b direction and exceeds $%
H_T$ in the vortex core. Then the magnetic subsystem in the core undergoes
the spin-flop transition.

2. The external magnetic field is applied along the a direction. Then the AF
alignment is unstable against spin-flop transformation and the vortex state
is created in the uniform spin-flop phase.

Assume the first case. The field intensity in the vortex core in the low
vortex density regime doubles the value of the external field intensity \cite
{10}. The vortices are created in the uniform antiferromagnetic medium if
the condition $Hc_1<H<\frac 12H_T$ is fulfilled. When the external field
reaches $\frac 12H_T$, the magnetic field inside the core is $H_T$ whereas
outside it decreases from this value. In this way the spin-flop domain
appears along the vortex. Finally, depending on the external field
intensity, there can be two types of vortices in the specimen. On the other
hand, in the case 2, the vortices do not change their structure and are
similar to that in nonmagnetic HTS.

The magnetic field aligned parallel to the conducting planes makes the
vortex lattice to accommodate itself to the layer structure so that the
vortex cores lie in between the superconducting sheets. A current density $j$
flowing along the planes exerts a Lorentz force on the vortices in the c
direction. Intrinsic pinning barriers are formed on strongly superconducting
layers. The vortices loose large amount of condensation energy when they
move across the barriers. Depending on the external field intensity and its
direction in the a-b plane we deal with vortices of two types. It is
interesting therefore to see how changes the intrinsic pinning barrier
during the creation of magnetic domain along the vortex.

\section{ LONDON EQUATIONS}

The electrodynamics of layered superconductors is based on the
Lawrence-Doniach model that in the London approximation gives the following
functional of the free energy of superconducting subsystem \cite{11}

\begin{equation}
F_s=\frac{\varphi _0^2d}{8\pi ^2\mu _0\lambda _{ab}^2}\sum\limits_n\int
\left\{ (\mathbf{\nabla }\Phi _n+\frac{2\pi }{\varphi _0}\mathbf{A}%
_p)^2+\frac 2{r_j^2}\left[ 1-\cos (\chi _n)\right] \right\} d^2r\text{,}
\label{1}
\end{equation}
where $r_j=d(\lambda _c/\lambda _{ab}),\mathbf{A}=(\mathbf{A}_p,A_z),\mathbf{%
A}_p=(A_x,A_y),$ $\varphi _0$ denotes the flux quantum, $\mu _0$ the
magnetic permeability of the vacuum, $\lambda _c,\xi _c$ and $\lambda
_{ab},\xi _{ab}$ the magnetic field penetration depth and the coherence
length parallel and perpendicular to the layers respectively. The parameter $%
r_j$ plays the role of the effective radius of the vortex core. The term
with cosine gives the Josephson current due to the gauge invariant phase
difference between the layers

\begin{equation}
\chi _n=\Phi _{n+1}-\Phi _n+\frac{2\pi }{\varphi _0}\int%
\limits_{(n+1)d}^{nd}A_zdz\text{.}  \label{2}
\end{equation}
The antiferromagnetic two sublattice subsystem with single ion anisotropy is
described with the following free energy functional:

\begin{equation}
F_m=d_i\sum\limits_n\int \left\{ J\mathbf{M}_{1n}\mathbf{M}%
_{2n}-K\sum\limits_{i=1}^2M_{in\perp }-\gamma \sum_{i=1}^2\sum_{\alpha
=x,y,z}(\mathbf{\nabla }M_{in\alpha })^2\right\} f_{n,n+1}(z)d^2r\text{,}
\label{3}
\end{equation}
where $\mathbf{M}_n=\mathbf{M}_{1n}+\mathbf{M}_{2n}$ is the sum of the
magnetization vectors of the sublattices in the n -th insulating layer, $%
M_{in\perp }$ is the component along the anisotropy axis of the
magnetization sublattice vector in the n-th layer, $J$ denotes the exchange
constant between two sublattices, $K$ is the single ion anisotropy constant, 
$\sqrt{\gamma }$ is the magnetic stiffness length, the factor $%
f_{n,n+1}(z)=1 $ for insulating layers and 0 otherwise, and\ $M=\left| 
\mathbf{M}_{1n}\right| =\left| \mathbf{M}_{2n}\right| $. Finally we add the
magnetic field energy to obtain the free energy of the entire system

\begin{equation}
F=F_s+F_m+\frac{\mu _0}2\int \mathbf{H}^2d^2rdz\text{.}  \label{4}
\end{equation}
The coupling between the magnetic and superconducting subsystem is supposed
to be electromagnetic. This means that both order parameters are coupled
through the vector potential $\mathbf{A}$

\begin{equation}
\mathbf{B}=\limfunc{rot}\mathbf{A}=\mu _0\mathbf{H}+\mathbf{M}\text{,}
\label{5}
\end{equation}

\begin{equation}
\left| \Psi \right| ^2\approx j_s=\left| \limfunc{rot}\mathbf{H}\right| 
\text{,}  \label{6}
\end{equation}
where $\mathbf{B}\ $( in MKSA units) is the vector of a magnetic flux
density ( magnetic induction) and $\mathbf{H}\ $ is the vector of a
thermodynamic magnetic field intensity. The analysis of a stability of the
functional (3) against the spin-flop transition gives the thermodynamical
critical field of the upper limit of the AF configuration equal to 
\[
H_T=M_0\sqrt{K(J-K)}\text{.} 
\]
The London equations resulting from (4) are following:

\begin{eqnarray}
&&B_x+\lambda _c^2\frac \partial {\partial y}\limfunc{rot}\nolimits_z(%
\mathbf{B}-\mathbf{M})-\lambda _{ab}^2\frac \partial {\partial z}\limfunc{rot%
}\nolimits_y(\mathbf{B}-\mathbf{M})=\varphi _0\delta (y)\delta (z)\text{,} 
\nonumber  \label{7} \\
&&B_y+\lambda _{ab}^2\frac \partial {\partial z}\limfunc{rot}\nolimits_x(%
\mathbf{B}-\mathbf{M})-\lambda _c^2\frac \partial {\partial x}\limfunc{rot}%
\nolimits_z(\mathbf{B}-\mathbf{M})=0\text{,}  \label{7} \\
&&B_z+\lambda _{ab}^2\frac \partial {\partial x}\limfunc{rot}\nolimits_y(%
\mathbf{B}-\mathbf{M})-\lambda _{ab}^2\frac \partial {\partial y}\limfunc{rot%
}\nolimits_x(\mathbf{B}-\mathbf{M})=0\text{.}  \nonumber
\end{eqnarray}
This equations should be supplemented by the appropriate set of differential
equations describing the spatial distribution of magnetization. Simple
conjecture, as shall be seen later, can make the calculations less complex
and at the same time does not oversimplify the problem. The London model of
continuous superconductors may be used at length scales larger than the
coherence length, i.e. the core dimension. The structure of a vortex lying
in the a-b plane in a layered superconductor with Josephson coupling between
adjacent layers resembles the Abrikosov`s one except that the order
parameter does not vanish anywhere \cite{12}. Instead there exists a region, 
$r_j$ along the plane and $d$ perpendicular to it, where the Josephson
current $j_z$ is of the order of the critical current. In this region, which
plays the role of the vortex core, the London model fails. Away from the
core the streamlines of the shielding supercurrents, which also represents
contours of constant magnetic field, are elliptical except for the zigzags
due to the intervening insulating layers(Fig.2). It is not crucial what kind
of transition is inherent in the magnetic subsystem: spin-flop, metamagnetic
or paramagnetic. The main fact is the appearance of a magnetic domain with
an induced magnetic moment $M$ along the vortex. We make the above mentioned
conjecture by assuming \cite{9} that the magnetic moment is constant across
the domain

\begin{equation}
\left| \mathbf{M}\right| =\left\{ 
\begin{array}{ccc}
M & if & \rho <\rho _m \\ 
0 & if & \rho >\rho _m
\end{array}
\right. \text{,}  \label{8}
\end{equation}
where $\ \rho _m$ is the dimensionless radius of the magnetic domain in the
coordinate system of the elliptical cylinder ($x=x,y=\lambda _c\rho \cos
\varphi ,z=\lambda _{ab}\rho \sin \varphi $).

Then, the set of equations (7) have the following solutions \cite{9}

\begin{eqnarray}
B_P &=&C_1I_0(\rho )+C_2K_0(\rho )~\text{ for }\rho <\rho _m\text{,}
\label{9} \\
B_{AF} &=&\frac{\mu _0H_T}{K_0(\rho _m)}K_0(\rho )\text{ ~for }\rho >\rho _m%
\text{,}  \nonumber
\end{eqnarray}
where $B_P$ and $B_{AF}$ denote the magnetic induction inside and outside
the domain respectively. The constants $C_1$ and $C_2$ are calculated using
(8) and the flux quantization condition to obtain

\begin{eqnarray}
C_1 &=&\frac{B_T\rho _mI_1(\rho _m)-D(\rho _m)I_0(\rho _m)}{1-I_0(\rho _m)}%
\text{,}  \nonumber  \label{10} \\
C_2 &=&\frac{D(\rho _m)K_0(\rho _m)+B_T\left[ \rho _mK_1(\rho _m)-1\right] }{%
1-I_0(\rho _m)}\text{,}  \label{10} \\
D(\rho _m) &=&\frac{\varphi _0}{2\pi \lambda _{ab}\lambda _c}-\mu _0H_T\frac{%
\rho _mK_1(\rho _m)}{K_0(\rho _m)}\text{,}  \nonumber
\end{eqnarray}
where $K_0$, $K_1$, $I_0$, $I_1$ denote the modified Bessel functions
whereas $\ B_T=M+\mu _0H_T$.\ Now we can think of $M$ as of experimental
parameter. To calculate $\rho _m$ one must first derive the line tension of
the vortex and then look for its minimum with respect to the domain radius.
Standard calculations give

\begin{equation}
\rho _m^2\approx \frac{5\varphi _0}{8\pi \lambda _{ab}\lambda _cB_T}\text{.}
\label{11}
\end{equation}
It is easy to see that almost all flux quantum is captured in the magnetic
domain.

\section{ ACTIVATION ENERGY}

The first quantitative approach toward intrinsic pinning in layered
superconductors was based on the observation that the superconducting order
parameter should have a periodic spatial variation across the layers \cite
{13}. For the present considerations, however, the method of critical
nucleus developed in \cite{14} is much more convenient. The activated
nucleus consists of a kink-antikink excitation, that is, a vortex line
segment is thrown to the adjacent layer, thereby creating two pancake
vortices of opposite ''vorticity'', as shown on Fig.3. The activation energy
can be regarded as the energy barrier for intrinsic pinning. Depending on
the magnitude of the driving current density the process may continue as the
single vortex activation or the activation of the vortex bundle. First,
consider the activation of a segment of a vortex to the neighboring
interlayer spacing (Fig.3). The energy associated with this process can be
written as:

\begin{equation}
U=\delta E+V_{K,-K}(R)-(j-j_0)\varphi _0dR.  \label{12}
\end{equation}
The subscript ''a'' or ''b'' (of $U$, $\delta E$ and $j_0$ ) indicates that
this quantity is calculated for a or b direction in the plane. $\delta E$ is
the amount of condensation and magnetic domain energy that is lost at two
points of the layer threaded by the kinks separated by a distance R (see
Fig.3). $V_{K,-K}(R)$ is the kink-antikink interaction energy. The term
proportional to the driving current $j$ is due to the Lorentz force. The
term proportional to $j_0$ is the energy associated with the distortion of
the line due to the formation of nucleus. This term can be estimated from
the simple considerations \cite{14} $\ $%
\[
j_0\varphi _0d\sim \frac 12\int dydzC(y,z)\left( \frac{\partial u_z}{%
\partial z}\right) ^2\text{,} 
\]
\ where the Fourier transform of the compression modulus is given by \cite
{15} 
\[
C(k_y,k_z)=\frac{B^2}{\mu _0(1+\lambda _{ab}^2k_z^2+\lambda _c^2k_y^2)}\text{
.} 
\]
\ By taking $dydz\sim \varphi _0/B,u_z\sim d,\frac \partial {\partial z}\sim
k_z\sim k_y(\lambda _c/\lambda _{ab})$ the integral can be estimated as
follows

\begin{equation}
j_{0a}=\frac{Bd}{4\lambda _{ab}^2}.  \label{13}
\end{equation}
In the b direction, however, we have an additional contribution from the
magnetic domain $dydz\sim 5\varphi _0/8B_T$ according to (11), so we get

\begin{equation}
j_{0b}=j_{0a}+\frac{5dB_T}{128\lambda _{ab}^2}\text{.}  \label{14}
\end{equation}
As the current density $j$ drops below $j_0$ a single-vortex line can no
longer be activated due to the confinement energy provided by the vortex
lattice. The energy $\delta E$ is calculated from (7) with the right-hand
sides representing the vortex cores:

\begin{eqnarray*}
&&\left\{ \left| x\right| >\frac R2,y=0,z=0\right\} \text{,}\left\{ \left|
x\right| <\frac R2,y=0,z=-d\right\} \text{ } \\
&&\text{and }\left\{ x=\pm \frac R2,y=0,0<z<-d\right\} \text{.}
\end{eqnarray*}
The solution is then substituted to the free energy functional (4). Taking
the limit $R\rightarrow \infty $ we exclude the energy of the kink-antikink
interaction. The calculations are involved, so we write down only the
results.

\begin{equation}
\delta E_a=2d\epsilon _0\ln \frac{r_j}{\xi _{ab}}\text{,}  \label{15}
\end{equation}

\begin{equation}
\delta E_b=d\epsilon _b\ln \frac{r_j}{\xi _{ab}}\text{ , ~}  \label{16}
\end{equation}
where $\epsilon _0=\varphi _0^2/(16\pi ^2\mu _0\lambda _{ab}^2),\epsilon _b=%
\frac{77}{64}\epsilon _0\ln \left[ \varphi _0/(\pi r_j^2B_T)\right] .$

The energy of kink-antikink interaction was calculated in \cite{14}

\begin{equation}
V_{K,-K}(R)=-\frac{d^2\epsilon _0}{2\lambda _{ab}}f(\frac R{\lambda _c})%
\text{ ,~}  \label{17}
\end{equation}

\[
\text{where ~}f(\frac R{\lambda _c})=\left\{ 
\begin{array}{ccc}
(\lambda _c/R)-\ln (r_j/\xi _{ab}) & \text{for} & 
\begin{array}{cc}
r_j<<R<< & \lambda _c
\end{array}
\\ 
2(\lambda _c/R)^3\exp \left( -R/\lambda _c\right) & \text{for} & R>>\lambda
_c
\end{array}
\right. \text{.} 
\]
We introduce the Ginzburg-Landau critical current density $j_{GL}=4\epsilon
_0/(\varphi _0\xi _{ab}3\sqrt{3})$ and the quantity $%
I_{a,b}=2(j-j_{0a,b})/(j_{GL}3\sqrt{3})$, then (12) can be rewritten in the
following way\ 
\begin{eqnarray}
U_a &=&2d\epsilon _0\left\{ \ln \frac{r_j}{\xi _{ab}}+I_a\frac R{\xi
_{ab}}-\frac d{4\lambda _{ab}}\text{~}f(\frac R{\lambda _c})\right\} \text{,}
\label{18} \\
U_b &=&d\left\{ \epsilon _b\ln \frac{r_j}{\xi _{ab}}+2\epsilon _0I_b\frac
R{\xi _{ab}}-\frac{d\epsilon _0}{2\lambda _{ab}}\text{~}f(\frac R{\lambda
_c})\right\} \text{.}  \nonumber
\end{eqnarray}
The critical size of the nucleus $\ R_c$ \ is given as a minimum of (18)
with respect to $R$. In the approximation\ \ $r_j<<R<<\lambda _c\ $ \
corresponding to the current regime $\xi _cd/\lambda _{ab}^2<<I_{a,b}<<\xi
_c/d\ $\ we get

\begin{eqnarray}
R_{ca,b}^2 &=&\xi _{ab}^2\frac d{4I_{a,b}\xi _c}\text{,}  \nonumber
\label{19} \\
U_a^c &=&2d\epsilon _0\left\{ \ln \left( \frac{r_j}{\xi _{ab}}\right) -\sqrt{%
\frac{dI_a}{\xi _c}}\right\} \text{,}  \label{19} \\
U_b^c &=&d\left\{ \epsilon _b\ln \left( \frac{r_j}{\xi _{ab}}\right)
-2\epsilon _0\sqrt{\frac{dI_b}{\xi _c}}\right\} \text{.}  \nonumber
\end{eqnarray}
For the opposite case $\ R>>\lambda _c$\ and$\ \xi _cd/\lambda
_{ab}^2>>I_{a,b}$

\begin{eqnarray}
R_{ca,b} &=&\lambda _c\ln \left( \frac{d\xi _c}{I_a\lambda _{ab}^2}\right) 
\text{,}  \nonumber \\
U_a^c &=&2d\epsilon _0\left\{ \ln \left( \frac{r_j}{\xi _{ab}}\right) -\frac{%
\lambda _{ab}I_a}{\xi _c}\ln \left( \frac{d\xi _c}{I_a\lambda _{ab}^2}%
\right) \right\} \text{,}  \label{20} \\
U_b^c &=&d\left\{ \epsilon _b\ln \left( \frac{r_j}{\xi _{ab}}\right) -\frac{%
\lambda _{ab}I_b}{\xi _c}\ln \left( \frac{d\xi _c}{I_b\lambda _{ab}^2}%
\right) \right\} \text{.}  \nonumber
\end{eqnarray}
When the driving current drops below\ $j_0\ $the critical nucleus is 3D
object (Fig.4). In our case it is a parallelepiped of the height $\ R\ $
along the bundle and of the section \ $S\ $ across it. The activation energy
is a sum of the volume energy due to the Lorentz force and the surface
energy.

\begin{equation}
U_{a,b}=-jBdRS+\delta E_{a,b}\left( \frac{BS}{\varphi _0}\right) +j_{0a,b}dR%
\sqrt{BS\varphi _0}\text{.}  \label{21}
\end{equation}
The second term is the loss of condensation energy (and magnetic domain
energy in the case of b direction) on both surfaces perpendicular to the
bundle multiplied by the number of vortices threading these surfaces. The
third term is the elastic energy released in the surface parallel to the
shifted vortex $\ j_{0a,b}dR\varphi _0\ $ multiplied by the number of
shifted vortices $\sqrt{BS/\varphi _0}$ (one vortex per plane). The critical
nucleus is then $S_c=(\varphi _0/B)(j_{0a,b}/j)^2$ , $R_c=\delta
E_{a,b}/(jd\varphi _0)$, and the activation energy is

\begin{equation}
U_{a,b}^c=\delta E_{a,b}\left( \frac{j_{0a,b}}j\right) ^2.  \label{22}
\end{equation}

\section{ MOTION OF THE FLUX}

The resistive mechanism in the mixed state is determined by the activation
process leading to magnetic flux motion (creep). This motion induces
electric field which can be observed on the current-voltage characteristic.
We consider the motion of activated kinks along the layers of the length $L$
along the magnetic field direction. Assume that each one can reach the
boundary of the sample before the new one is created. The mean electric
field associated with this motion is given by

\begin{equation}
E=BPLdS_c\text{ ,}  \label{23}
\end{equation}
where $P$ is the activation probability per unit volume and unit time. For
thermal activation this probability is given by $\ P=\alpha \exp \left(
-U_c/k_BT\right) \ $. There is however a crossover temperature $T_0$ \cite
{16} below which quantum tunneling of vortices is dominating. The
probability for quantum tunneling is given by $\ P=\beta \exp \left(
-U_c/k_BT_0\right) $ and remains finite even for $T=0$. The temperature $T_0$
depends on the driving current, critical current, upper critical field, and
resistivity in the normal state. For our purposes it is worth to note that
the quantity\ $\exp \left( -U_c/k_BT_0\right) \ $ will contribute only for
driving currents of the order of the critical current \cite{16}. The Neel
temperature for layered antiferromagnetic superconductors varies from
hundreds of mK (0.6K for ErBa$_2$Cu$_3$O$_7$) to several Kelvin (6.8K for
ErNi$_2$B$_2$C) and therefore both mechanisms of activation are present in
these compounds. The preexponential factors and $T_0$ cannot be calculated
in the framework of thermodynamic considerations alone. Fortunately, it was
shown in \cite{17} that the activation probability of macroscopic quantum
excitations is proportional to $j^3$. Thus we can assume that $P=\alpha
_0j^3\exp \left( -U_c/k_BT\right) $. Now we can calculate the
current-voltage characteristics for the current density regimes considered
previously. For $j<<j_0$

\begin{eqnarray}
E_a &=&\varphi _0dL\alpha _0j_{0a}^2j\exp \left\{ -\frac{\delta E_a}{k_BT}%
\left( \frac{j_{0a}}j\right) ^2\right\} \text{,}  \label{24} \\
E_b &=&\varphi _0dL\alpha _0j_{0b}^2j\exp \left\{ -\frac{\delta E_b}{k_BT}%
\left( \frac{j_{0b}}j\right) ^2\right\} \text{.}  \nonumber
\end{eqnarray}
This almost linear dependence of $E$ on $j$ indicates that the resistive
mechanism of bundle activation follows Ohm law.

For $j>>j_0\ $and $\ \xi _cd/\lambda _{ab}^2<<I_{a,b}<<\xi _c/d\ $

\begin{eqnarray}
E_a &=&\varphi _0dL\alpha _0j^3\exp \left\{ -\frac{\delta E_a}{k_BT}+\frac{%
2d\epsilon _0}{k_BT}\sqrt{\frac{dI_a}{\xi _c}}\right\} \text{,}  \label{25}
\\
E_b &=&\varphi _0dL\alpha _0j^3\exp \left\{ -\frac{\delta E_b}{k_BT}+\frac{%
2d\epsilon _0}{k_BT}\sqrt{\frac{dI_b}{\xi _c}}\right\} \text{ .}  \nonumber
\end{eqnarray}
For $j>>j_0\ $and$\ \xi _cd/\lambda _{ab}^2>>I_{a,b}$

\begin{eqnarray}
E_a &=&\varphi _0dL\alpha _0j^3\exp \left\{ -\frac{\delta E_a}{k_BT}+\frac{%
2d\epsilon _0}{k_BT}\frac{\lambda _{ab}I_a}{\xi _c}\ln \left( \frac{d\xi _c}{%
\lambda _{ab}^2I_a}\right) \right\} \text{,}  \label{26} \\
E_b &=&\varphi _0dL\alpha _0j^3\exp \left\{ -\frac{\delta E_b}{k_BT}+\frac{%
2d\epsilon _0}{k_BT}\frac{\lambda _{ab}I_b}{\xi _c}\ln \left( \frac{d\xi _c}{%
\lambda _{ab}^2I_b}\right) \right\} \text{.}  \nonumber
\end{eqnarray}
We can also calculate the rate of flux creep due to the thermal activation
of vortices. To do this consider hollow cylindrical sample of a radius $r$
and the wall thickness\ $l<<r$\ placed in the magnetic field $\
B_{ex}>B_{c1}\ $ applied parallel to the cylinder axis. The sample has the
trapped field \ $B_{in}\ $ inside the hole and trapped flux $\ \Phi
=(B_{in}-B_{ex})\pi r^2$. According to the Faraday's law electric field due
to the change of the trapped flux is equal to $\ (\mu _0/2)lr(dj/dt)$.
Combining this result with (23) we have finally

\begin{equation}
BPLdS_c+\ \frac 12\mu _0lr\frac{dj}{dt}=0\text{ .}  \label{27}
\end{equation}

This equation can be solved analytically only in the case of the weak
currents. Consequently for excitations in the form of bundle of vortices
(27) is written as

\begin{equation}
\Omega j\exp \left\{ -\frac{\delta E_{a,b}}{k_bT}\left( \frac{j_{0a,b}}%
j\right) ^2\right\} +\frac{dj}{dt}=0\text{ ,}  \label{28}
\end{equation}
where $\Omega =\varphi _0\alpha _0j_0^2/(\mu _0\gamma )$ and $\gamma
=rl/(Ld) $ is the factor determined by the geometry of the sample. The
solution of (28) is given in terms of exponential integrals and for the case
of $j_{0a,b}/j\ -1<<1\ $it can be approximated as:

\begin{equation}
\frac{j(0)}{j(t)}-1=\frac{\Phi (0)}{\Phi (t)}-1=\frac{k_BT}{2\delta E_{a,b}}%
\left( \frac{j(0)}{j_{0a,b}}\right) ^2\ln \left( 1+\omega _{a,b}t\right) 
\text{,}  \label{29}
\end{equation}
where

\begin{equation}
\omega _{a,b}=\frac{4\varphi _0\alpha _0\delta E_{a,b}}{\mu _0\gamma }\left( 
\frac{j_{0a,b}}{j(0)}\right) \exp \left\{ -\frac{\delta E_{a,b}}{k_bT}\left( 
\frac{j_{0a,b}}{j(0)}\right) ^2\right\} \text{ .}  \label{30}
\end{equation}
This result is in agreement with the experiments on nonmagnetic HTSC ( see 
\cite{18}). For $\ 0<<t<<1/\omega \ $ the change of trapped flux is linear
in time and for\ $t>>1/\omega \ $ logarithmic. In antiferromagnetic
superconductors, however, we see additional change of characteristic
frequency as the magnetic field changes its direction in the a-b plane.

\section{ CONCLUSIONS}

The current\ $j_0$\ plays very important role in all above calculations. For
the following values $\ d\sim 10^{-9}m$, $\xi _c\sim 3\times 10^{-10}m$, $%
\xi _{ab}\sim 3\times 10^{-9}m$, $r_j\sim 10^{-8}m\ $ and their typical
temperature dependence we can estimate $\ $%
\[
\frac{j_{0a}}{j_{GL}}\sim \frac{Bd\xi _{ab}}{\varphi _0}\sim 10^{-3}B\sqrt{%
\frac{T_c}{T_c-T}\ }\text{.} 
\]
Since the depairing current is of the order of $\ 10^{13}\left[
T_c/(T_c-T)\right] ^{3/2}\ $ then the current\ $j_{0a}$\ is of the order of\ 
$10^{10}B\left[ T_c/(T_c-T)\right] \ .$ Although there are no precise
measurements of spin-flop transition in the antiferromagnetic high
temperature superconductors, we assume that $\ \mu _0H_T\sim 40mT\ $. The
typical value of $\ 5.5\mu _B$\ per rare-earth atom per unit cell gives $\
M\sim 0.37T$. It is possible now to estimate the change of\ $j_0\ $ due to
the creation of spin-flop domain along the vortex :

\[
\ \frac{j_{0a}}{j_{0b}}\sim 1+0.625\frac{B_T}B\sim 3 
\]
\ providing that\ $H_T/H_{c1}-1<<1$. Similar considerations show that the
characteristic frequency of flux creep changes of order of magnitude when
the spin-flop domain is created.

\section{ACKNOWLEDGEMENTS}

This work was financially supported by Komitet Badan Naukowych under grant
2PO3B17109.

\newpage

\textbf{FIGURE CAPTIONS}

\textbf{Fig.1.} Schematic drawing of a piece of a layered superconductor.
Hatched areas (n,n+1) represent superconducting layers with thickness d$_s$.
Bold arrows represent magnetic moments of RE ions lying in the isolating
layers of thickness d$_i$. The reference frame and the crystallographic axes
are shown.

\textbf{Fig.2.} Single vortex line lying in the a-b plane and the induced
spin-flop domain along its core.

\textbf{Fig.3.} Single vortex activation.

\textbf{Fig.4.} Activated bundle of vortices.


\newpage

\begin{thebibliography}{99}
\bibitem{1}  {\O }. Fischer and M. B. Maple, eds. Superconductivity in
ternary compounds (Springer, Berlin, 1983).

\bibitem{2}  J. W. Lynn, J. Alloys and Compounds \textbf{181 }(1992) 419.

\bibitem{3}  J. W.Lynn, T. W. Clinton, W-H. Li, R. W. Erwin, J. Z .Lin, R.
N. Shelton and P.Klavins, J. Appl. Phys., \textbf{67 }(1990) 4533.

\bibitem{4}  J. Zaretsky, C. Stassis, A. I. Goldman, P. C. Canfield, P.
Dervenagas, B. K. Cho and D. C. Johnston, Phys. Rev., \textbf{B51} (1995)
678.

\bibitem{5}  S. K. Sinha, J. W. Lynn, T. E. Grigereit, Z. Hossain, L. C.
Gupta, R. Nagarajan, C. Godard, Phys.Rev., \textbf{B51} (1995) 681.

\bibitem{6}  A. Morrish, The physical principles of magnetism (J. Wiley, New
York 1965).

\bibitem{7}  T. Krzyszto\'{n}, J. Magn. Magn. Materials, \textbf{15-18 }%
(1980) 1572 and Phys. Lett., \textbf{104A }(1984) 225.

\bibitem{8}  W. Thomlinson, G. Shirane, D. E. Moncton, M. Ishikawa, {\O }.
Fischer, J. Appl. Phys., \textbf{50 }(1979) 1981 and Phys. Rev., \textbf{B23}
(1981) 4455

\bibitem{9}  T. Krzyszto\'{n}, Phys.Lett., \textbf{A190} (1994) 196.

\bibitem{10}  M. Tinkham, Introduction to superconductivity, chapter 5, eqn
(5-37), (McGraw-Hill Inc.,New York,1975).

\bibitem{11}  L. N. Bulaevskii, M. Ledvij and V. G. Kogan, Phys. Rev.,%
\textbf{\ B46} (1992) 366.

\bibitem{12}  J. R. Clem and M. W. Coffey, Phys. Rev., \textbf{B42} (1990)
6209.

\bibitem{13}  M. Tachiki, S. Takahasi, Solid State Commun., \textbf{70 }%
(1989) 291.

\bibitem{14}  S. Chakravarty, B. I. Ivlev, Y. N. Ovchinnikov, Phys. Rev.,%
\textbf{\ B42} (1990) 2143.

\bibitem{15}  E. H. Brandt, Physica, \textbf{C195} (1992) 1.

\bibitem{16}  B. I. Ivlev, Y. N. Ovchinnikov, R. S. Thompson, Phys. Rev., 
\textbf{B44} (1991) 7023.

\bibitem{17}  P. Tekiel, to be published in Z.Phys.B

\bibitem{18}  G. Blatter, M. V. Feigelman, V. B. Geshkenbein, A. I. Larkin,
Rev. Mod. Phys, \textbf{66} (1994)1125.
\end{thebibliography}
\end{document}